\documentclass[12pt]{article}

\usepackage{amsmath, cite, graphicx, subfigure}

 \numberwithin{equation}{section}

\begin{document}  

\title{\bf Polymer Adsorption on Curved Surfaces:
Finite chain length corrections}
\author{K.I. Skau and E.M. Blokhuis\\ 
Colloid and Interface Science, Leiden Institute of Chemistry,\\
Gorlaeus Laboratory, P.O. Box 9502, 2300 RA Leiden,\\
The Netherlands}
\date{March 18, 2003}
\maketitle


\begin{abstract}

\noindent
The structural properties of polymers adsorbed onto a surface have
been widely investigated using self-consistent mean-field theories.
Recently, analytical mean-field theories have been applied to study
polymer adsorption on curved surfaces but all in the context of the
ground state dominance approximation in which the polymer chain length
($N$) is essentially infinite.  Using an expression for the free
energy by Semenov, we determine leading order (in $1/N$) corrections
due to the finiteness of the polymer chain length on surface tension,
spontaneous curvature, and rigidity constants.

\end{abstract}

\pagebreak

\section{Introduction}

The influence of polymer adsorption onto surfaces has been a topic
that has attracted much attention for several decades. This is not
only because of its great practical interest, but also because it
serves as a testing ground of theoretical models for confined polymer
systems \cite{de_Gennes, Fleer_book, Eisenriegler_book}.  Recently,
investigations have been dedicated to study the interplay between
colloidal particles and polymers \cite{Odijk, Johner_01, Lekkerkerker,
Eisenriegler_Dietrich, Maasen, Louis}.  In these systems the
\emph{curvature} of the surface of the colloidal particle becomes a
factor, and work has focused on the situation where either the
colloidal particle is much larger or much smaller than the polymer
coil \cite{Odijk, Johner_01}.  When the radius of curvature is large,
it is reasonable to expand the free energy for polymer adsorption in
curvature.  Helfrich supplied the general form of the surface free
energy expanded to second order in the curvature
\cite{Helfrich_Canham}:
\begin{equation}
F_{H} = \int\!\! dA \, \biggl[ \sigma  - \frac{2 k}{R_0} \, \bigl(\frac{1}{R_1}
+ \frac{1}{R_2} \bigr) + \frac{k}{2} \, \bigl(\frac{1}{R_1}
+ \frac{1}{R_2} \bigr)^2 + \bar{k} \frac{1}{R_1 R_2} \biggr] \,, 
\label{eq:Helfrich_free_energy}
\end{equation}
where $R_1$ and $R_2$ are the local radii of curvature. Apart from the
surface tension, $\sigma$, one can identify three parameters in the
Helfrich free energy that describe the physical properties of the
curved interface: the spontaneous curvature $1/R_0$, the bending
rigidity constant $k$, and the Gaussian rigidity $\bar{k}$.

A great deal of attention has turned to the determination of the value
of these curvature parameters in complex systems, both theoretically
and experimentally. For surfaces interacting with polymers, the
curvature parameters were calculated by a number of groups
\cite{Eisenriegler_book, Eisenriegler_Dietrich, Maasen, Louis,
Ji_Hone, Brooks_Marques_Cates, Podgornik, Clement_Joanny, Skau}. In
the work by Eisenriegler and others \cite{Eisenriegler_book,
Eisenriegler_Dietrich, Maasen, Louis}, the polymer density is assumed
to be zero at the surface of the colloidal particle so that a
depletion layer exists around each particle.  For surfaces with
\emph{enhanced} polymer adsorption, calculations
\cite{Brooks_Marques_Cates, Clement_Joanny, Skau} were performed in
the context of the ground state dominance approximation
\cite{de_Gennes, Edwards, Lifshitz} in which the polymer chain length
is essentially infinite. In the present work, we use recent extensions
to ground state dominance to determine corrections to the curvature
parameters due to the finite length of the polymer chain. Such a
calculation is of interest since the polymer chain length is an
important parameter in experiments and computer simulations
\cite{Simulations} thus providing a more stringent testing of
theoretical models. Moreover, it was expected \cite{Fleer_book,
Semenov_95}, and later verified \cite{Semenov_96, Johner_96, Bonet_96,
Semenov_96_review, Semenov_98}, that in certain situations, the
`tails' of the polymer chain become important, leading to
\emph{qualitatively} different behavior. An example is the repulsion
between two planar walls when the separation is of the order of the
polymer's radius of gyration \cite{Bonet_96}. In the ground state
dominance model the radius of gyration is infinite and the interaction
between plates is attractive for all separations \cite{Bonet_96,
de_Gennes_82}.

The first extension of the ground state dominance model was proposed
by Semenov, Bonet-Avalos, Johner, and Joanny \cite{Semenov_96}.  They
took the presence of tails into account by including a \emph{second}
order parameter related to the end segment density. Good agreement was
obtained for the loop and tail distribution of adsorbed polymer when
the theoretical predictions are compared with lattice self-consistent
mean-field calculations \cite{Johner_96}.  The two-order parameter
model was later extended to also take into account the presence of
free (non-adsorbed) polymer away from the surface \cite{Johner_96,
Semenov_96_review, Semenov_98}. Subsequently, Semenov showed that the
two-order parameter model may be cast into a free energy formalism of
a single order parameter \cite{Semenov_96_review} in which the
Euler-Lagrange equations are the Edwards equations, just as de Gennes
had previously done in the context of the ground state dominance
approximation \cite{de_Gennes}.  In this article, we use the
expression for the free energy by Semenov \cite{Semenov_96_review} to
determine the polymer length corrections to the curvature parameters.

The outline of the paper is as follows: In the next section we review
the description of finite chain length corrections to polymer
adsorption on a planar wall in the two-order parameter model. We show
how from the two-order parameter model a \emph{single} order parameter
model, to which we refer to as the Semenov model, is constructed to
describe leading order corrections in $1/N$ to ground state
dominance. In section 3, the single order parameter model is applied
to the study of curved surfaces.  Explicit expressions for the surface
tension and curvature parameters are given. We end with a discussion
of results.

\section{Semenov model for adsorption onto a planar surface}

Before addressing the properties of curved surfaces, we discuss finite
chain length corrections to the structure and tension of polymer
adsorption onto a \emph{planar} surface.  We then generalize the
description to polymer adsorption onto curved surfaces.

The situation under consideration is that of a polymer chain adsorbed
onto a planar wall located at $z\!=\!0$ (see Figure 1).  The
calculations presented here are performed in the context of
self-consistent mean-field theory whereby a \emph{single} polymer chain
is considered with the effect of the other polymers taken into account
by the presence of an external field proportional to the local segment
density \cite{Edwards}.  Furthermore, the interaction of the polymer
with the wall is taken into account through an infinitely short-ranged
(attractive) interaction potential \cite{de_Gennes}.

Before taking the finite length of the polymer chain into account, we
remind ourselves of the de Gennes-Lifshitz description
\cite{de_Gennes, Lifshitz} for polymer adsorption of an \emph{infinite}
chain, the so-called \emph{ground state dominance approximation}. The
free energy is a functional of the polymer segment density $\phi(z)$
\cite{de_Gennes}:
\begin{equation}
\frac{F[\phi]}{A \, k_{\rm B} T} = \int\limits_{0}^{\infty} \!\! dz
\left[ \frac{a^2}{24} \, \frac{\phi^{\prime}(z)^2}{\phi(z)} +
\frac{v}{2} \, [\phi(z) - \phi_b]^2 \right] - \frac{1}{d} \,
\frac{a^2}{6} \, \phi_w \,,
\label{eq:GSD_free_energy_phi}
\end{equation}
where $k_{\rm B}$ is Boltzmann's constant, $T$ is the temperature, $A$
the surface area, and $a$ is the polymer segment length.  The
subscripts $b$ and $w$ refer to the value of the polymer segment
density in the bulk, $\phi_b\!\equiv\!\lim\limits_{z\rightarrow\infty}
\phi(z)$, and at the wall, $\phi_w\!\equiv\!\phi(0)$, respectively.
The first term is the Lifshitz expression for the chain entropy
\cite{Lifshitz}, the second term gives the mean-field interaction
between polymer segments in a good solvent ($v$ is the
\emph{excluded volume} parameter), and the last term accounts for
the polymer interaction with the surface \cite{de_Gennes} ($d$ is the
\emph{extrapolation length}; its inverse measures the interaction
strength with the surface).

The above free energy is the \emph{excess} or \emph{surface} free
energy since it is constructed such that when the polymer segment
density becomes equal to the bulk segment density,
$\phi(z)\!\rightarrow\!\phi_b$, the integrand reduces to zero. The
result is that the surface tension is equal to the above free energy
per area when evaluated at its minimum, i.e.~when the `optimum'
segment density profile, determined by the Euler-Lagrange equation, is
inserted into it \cite{RW}.

Rather than the segment concentration, it is convenient to
consider the free energy as a functional of 
$\psi(z)\!\equiv\!\phi(z)^{1/2}$: 
\begin{equation}
\frac{F[\psi]}{A \, k_{\rm B} T} = 
\int\limits_{0}^{\infty} \!\! dz \left[
\frac{a^2}{6} \, \psi^{\prime}(z)^2
+ \frac{v}{2} \, [\psi(z)^2 - \phi_b]^2 \right]
- \frac{1}{d} \, \frac{a^2}{6} \, \psi_w^2 \,,
\label{eq:GSD_free_energy_psi}
\end{equation}
where $\psi_w\!\equiv\!\psi(0)$.
Minimization of the above free energy gives the following
Euler-Lagrange equation for $\psi(z)$ with boundary condition:
\begin{eqnarray}
\frac{a^2}{6} \, \psi^{\prime \prime}(z) &=&
v \, [ \psi(z)^2 - \phi_b] \, \psi(z) \,, \nonumber \\
\psi^{\prime}_w &=& - \frac{1}{d} \, \psi_w \,.
\label{eq:GSD_EL}
\end{eqnarray}
The Euler-Lagrange equation can be recognized as the Edwards equation
for an infinitely long polymer chain \cite{de_Gennes, Edwards}.

The expression for the free energy in eq \ref{eq:GSD_free_energy_psi}
has been the starting point of many density functional treatments of polymer
solutions \cite{de_Gennes, Fleer_book}. Owing to the nature of the ground state
dominance approximation, this treatment is, however, limited to infinitely long
chains. Recent investigations \cite{Semenov_96, Johner_96, Bonet_96,
Semenov_96_review, Semenov_98} have therefore been addressed to include
finite length corrections to the ground state dominance model. 

In order to account for the finiteness of the polymer chain length, a
second order parameter, $\varphi(z)$, was introduced \cite{Semenov_96}
connected to the description of chain \emph{ends}.
To understand better the physical origin of this term, consider
a polymer chain with at least one segment touching the
surface. Such a chain consists of a series of `loops'
with a `tail' on both sides \cite{Fleer_book} (see Figure 1).
The segment density of adsorbed chains, $\phi(z)$ is then a sum
of two terms; $\phi_{\ell}(z)\!=\!\psi(z)^2$ connected to the loop
segment density and $\phi_t(z)\!=\!\psi(z) \, \varphi(z)$ connected
to the tail segment density:
\begin{equation}
\phi(z) = \phi_{\ell}(z) + \phi_t(z) = \psi(z)^2 + \psi(z) \, \varphi(z) \,.
\label{eq:phi}
\end{equation}
This description is restricted to polymer chains that have
one or more segments touching the surface. As a consequence,
since the polymer chains are finite,
$\lim\limits_{z\rightarrow\infty} \phi(z)\!=\!0$.
The segment density $\phi(z)$ therefore does not describe the
polymers in the bulk region.

The density of \emph{chain ends}, $\phi_e(z)$, is proportional to
a single $\psi(z)$ \cite{Semenov_96}:
\begin{equation}
\phi_e(z) = B \, \psi(z) \,.
\label{eq:phi_e}
\end{equation}
The value of the proportionality constant $B$ is obtained by a normalization
onto the number of end-segments. To show this in more detail, we denote $\Gamma$ 
as the total adsorption of chains having one or more segments touching the wall:
\begin{equation}
\Gamma \equiv \int\limits_{0}^{\infty} \!\! dz \,
[\psi(z)^2 + \psi(z) \, \varphi(z) ] \,.
\label{eq:Gamma}
\end{equation}
The proportionality constant $B$ is then determined by the fact
that every adsorbed chain has \emph{two} ends so that the fraction
of end-segments therefore is $2/N$, with $N$ the total number
of segments \cite{Semenov_96}:
\begin{equation}
\int\limits_{0}^{\infty} \!\! dz \, B \, \psi(z) = \frac{2}{N} \, \Gamma
= \frac{2}{N} \, \int\limits_{0}^{\infty} \!\! dz \,
[ \psi(z)^2 + \psi(z) \, \varphi(z) ] \,.
\label{eq:B_1}
\end{equation}

The profiles of the two order parameters, $\psi(z)$ and $\varphi(z)$,
are determined by the following set of Edwards-like equations \cite{Semenov_96}:
\begin{eqnarray}
\frac{a^2}{6} \, \psi^{\prime \prime}(z) &=&
v \, [\psi(z)^2 + \psi(z) \, \varphi(z) + \epsilon - \phi_b] \, \psi(z) \nonumber \\
\frac{a^2}{6} \, \varphi^{\prime \prime}(z) &=&
v \, [\psi(z)^2 + \psi(z) \, \varphi(z) + \epsilon - \phi_b] \, \varphi(z) - B \,.
\label{eq:two-order_EL}
\end{eqnarray}
As boundary conditions at the surface, we have:
\begin{equation}
\psi^{\prime}_w =  - \frac{1}{d} \, \psi_w \,, \hspace*{50pt}
\varphi_w = 0 \,.
\label{eq:two-order_BC_wall}
\end{equation}
The latter condition follows from the interpretation of
$\psi(z) \, \varphi(z)$ as the tail segment density distribution,
that is zero at the surface by definition. As boundary conditions
far away from the surface, we have:
\begin{equation}
\lim\limits_{z\rightarrow\infty} \psi(z)\!=\!0 \,, \hspace*{50pt}
\lim\limits_{z\rightarrow\infty} \varphi(z)
= \frac{B}{v \, (\epsilon - \phi_b)} \,.
\label{eq:two-order_BC_bulk}
\end{equation}
The following functional is derived as a first integral of the
two differential equations in eq \ref{eq:two-order_EL} identified
as the free energy:
\begin{eqnarray}
\frac{F}{A \, k_{\rm B} T} &=& 
\int\limits_{0}^{\infty} \!\! dz \left[
\frac{a^2}{6} \, (\psi^{\prime})^2 + \frac{a^2}{6} \, \psi^{\prime} \,
\varphi^{\prime} + \frac{v}{2} \, [\psi^2 + \psi \, \varphi - \phi_b]^2
- B \psi \right. \nonumber \\
&& \hspace*{30pt} \left. + \epsilon \, (\psi^2 + \psi \, \varphi) \right]
- \frac{1}{d} \, \frac{a^2}{6} \, \psi_w^2 \,.
\label{eq:two-order_free_energy_1}
\end{eqnarray}
(For notational brevity we suppress the $z$-dependence of
$\psi$, and $\varphi$ in the free energy.)

The parameter $\epsilon$, which appears in the differential
equations for $\psi(z)$ and $\varphi(z)$, may be interpreted as the
chemical potential (in units of $k_{\rm B} T$) of the attached polymers.
Its value is determined by a balance of the chemical potential
of the polymers at the surface with the chemical potential of bulk polymer.
It is derived in ref.\cite{Semenov_96, Johner_96} that $\epsilon$ is
approximately given by:
\begin{equation}
\epsilon \simeq \frac{1}{N} \, \ln \, \biggl( \frac{\Gamma^2}
{\phi_b \, [ \int\limits_{0}^{\infty} \!\! dz \, \psi(z) ]^2} \biggr)  \,.
\label{eq:epsilon}
\end{equation}
Combining the above expression for $\epsilon$ with eq \ref{eq:B_1},
we can rewrite the normalization condition for $B$ as:
\begin{equation}
B \simeq \frac{2}{N} \, \phi_b^{1/2} \, e^{\epsilon \, N/2} \,.
\label{eq:B_2}
\end{equation}
This model is the two-order parameter model which has been shown
to be in agreement with numerical solutions of the self-consistent
field equations \cite{Johner_96}. One disadvantage of the two-order parameter
model is the rather artificial distinction made between chains that have one or
more segments touching the wall and free polymer chains, leading
to the presence of $\epsilon\!\neq\!0$.
Furthermore, the above formalism to determine the segment density
profiles and free energy does not correspond to a \emph{free
energy functional formalism} in which the differential equations
determining $\psi(z)$ and $\varphi(z)$ are the Euler-Lagrange equations
to the free energy in eq \ref{eq:two-order_free_energy_1}.
The reason is that a minimization of $F$ would not necessarily lead
to the boundary condition $\varphi(0)\!=\!0$, and, more importantly,
$B$ and $\epsilon$ are not constants in the minimization but are
themselves functionals of $\psi(z)$ and $\varphi(z)$
(eqs \ref{eq:B_1} and \ref{eq:epsilon}).

To circumvent these difficulties, we now turn to the derivation of
what we term the \emph{Semenov model} \cite{Semenov_96_review}.
In this model, no explicit distinction between adsorbed chains and free
chains is made leading to a \emph{single} order parameter describing the
total segment density, $\phi(z)$. As a first step we consider again the
free energy given by eq \ref{eq:two-order_free_energy_1}
\begin{equation}
\frac{F}{A \, k_{\rm B} T} = 
\int\limits_{0}^{\infty} \!\! dz \left[
\frac{a^2}{6} \, (\psi^{\prime})^2 +
\frac{a^2}{6} \, \psi^{\prime} \, \varphi^{\prime}
+ \frac{v}{2} \, [\psi^2 + \psi \, \varphi]^2 - B \psi \right]
- \frac{1}{d} \, \frac{a^2}{6} \, \psi_w^2 \,.
\label{eq:two-order_free_energy_2}
\end{equation}
Since adsorbed and free chains are now treated equally, we have
taken $\epsilon\!=\!0$. Furthermore, we have set $\phi_b\!=\!0$,
for the time being, and deal with the inclusion of bulk polymer below.

Next, since the combination $\psi(z) \, \varphi(z)$ is related to the
tail segment density, whose contribution to the total segment density
profile is of order $1/N$, we can conclude that $\varphi(z)$ describes
effects that are of order $1/N$, i.e.~$\varphi(z)\!=\!{\cal O}(1/N)$.
With $\phi(z)\!=\!\psi(z)^2 + \psi(z) \, \varphi(z)$, the total
segment density, we can therefore write in an expansion in $1/N$:
\begin{eqnarray}
\frac{1}{4} \frac{\phi^{\prime}(z)^2}{\phi(z)} &=& 
\psi^{\prime}(z)^2 + \psi^{\prime}(z) \, \varphi^{\prime}(z)
+ {\cal O} \left( \frac{1}{N^2} \right) \,, \nonumber \\
\phi(z)^{1/2} &=& \psi(z) + {\cal O} \left( \frac{1}{N} \right) \,.
\label{eq:expansion_phi_to_psi}
\end{eqnarray}
Inserting this into the free energy in eq \ref{eq:two-order_free_energy_2},
we have
\begin{equation}
\frac{F}{A \, k_{\rm B} T} = 
\int\limits_{0}^{\infty} \!\! dz \left[
\frac{a^2}{24} \, \frac{(\phi^{\prime})^2}{\phi}
+ \frac{v}{2} \, \phi^2 - B \, \phi^{1/2} \, \right]
- \frac{1}{d} \, \frac{a^2}{6} \, \phi_w \,,
\label{eq:Semenov_1}
\end{equation}
with corrections of ${\cal O}(1/N^2)$.
Next, we insert the explicit formula for $B$ given in
eq \ref{eq:B_2}, with $\epsilon\!=\!0$, into the
free energy. We are then left with the following
free energy functional:
\begin{equation}
\frac{F[\phi]}{A \, k_{\rm B} T} = 
\int\limits_{0}^{\infty} \!\! dz \left[
\frac{a^2}{24} \, \frac{(\phi^{\prime})^2}{\phi} + G(\phi) \, \right]
- \frac{1}{d} \, \frac{a^2}{6} \, \phi_w \,,
\label{eq:Semenov_2}
\end{equation}
where
\begin{equation}
G(\phi) = \frac{v}{2} \, \phi^2 - \frac{2}{N} \, \phi_b^{1/2} \, \phi^{1/2} \,.
\label{eq:G_phi_1}
\end{equation}
To impose a certain value of the bulk polymer density, we introduce the
bulk chemical potential in $G(\phi)$:
\begin{equation}
G(\phi) = \frac{v}{2} \, \phi^2 - \frac{2}{N} \, \phi_b^{1/2} \, \phi^{1/2}
- \mu \, \phi \,.
\label{eq:G_phi_2}
\end{equation}
The value of the bulk chemical potential, $\mu$, is chosen such that the
bulk density, as given by the \emph{minimum} of $G(\phi)$, is equal to
a certain $\phi_b$:
\begin{equation}
G^{\prime}(\phi_b) = v \, \phi_b - \frac{1}{N} - \mu = 0 
\hspace*{20pt} \Longrightarrow \hspace*{20pt}
\mu = v \, \phi_b - \frac{1}{N} \,.
\label{eq:G_phi_3}
\end{equation}
As expected, the chemical potential is lowered by reducing the chain length.
As a final step, we subtract from $G(\phi)$ the asymptotic bulk
free energy $G(\phi_b)$ in order for the above free energy to be
the surface free energy. Our final expression for $G(\phi)$ then reads:
\begin{equation}
G(\phi) = \frac{v}{2} \, (\phi - \phi_b)^2 
+ \frac{1}{N} \, (\phi^{1/2} - \phi_b^{1/2} )^2 \,.
\label{eq:G_phi_4}
\end{equation}
One notices that the above form of $G(\phi)$ reduces to that
in \emph{ground state dominance} (See eq \ref{eq:GSD_free_energy_phi})
in the limit $N\!\rightarrow\!\infty$, as it should.

Even though we started from an expression for the free energy in
eq \ref{eq:two-order_free_energy_1}, which we stressed is not the
free energy functional, we have now constructed a free energy
functional in eq \ref{eq:Semenov_2}. It is a functional of
$\phi(z)$, that is now the \emph{total} segment density profile in
which no reference to `loops' or `tails' is explicitly made.  Its form
was first proposed by Semenov \cite{Semenov_96_review} who also
determined the \emph{next} order contribution and showed it to be of
${\cal O}(1/N^{3/2})$.  Semenov \cite{Semenov_96_review} points out
that this approach agrees with the two-order parameter model (to
${\cal O}(1/N)$), even though this agreement has to be interpreted
somewhat loosely since a direct comparison is difficult to make owing
to the difference in the treatment of bulk polymer.

It is convenient to rescale all densities by $\phi_b$ and all lengths
by the bulk correlation length $\xi_b\!\equiv\!a/\sqrt{3 \, v \,
\phi_b}$:
\begin{equation}
x \equiv z / \xi_b \,, \hspace*{30pt}
\tilde{d} \equiv d / \xi_b \,, \hspace*{30pt}
\phi(x) \equiv \phi_b \, f_0(x)^2 \,.
\label{eq:rescaling}
\end{equation}
Furthermore, we introduce as a small parameter the (square of) the bulk
correlation length, $\xi_b$, divided by $R_G\!\equiv\!\sqrt{N a^2/6}$,
the polymer's radius of gyration:
\begin{equation}
\varepsilon \equiv \frac{\xi_b^2}{R_G^2} = \frac{2}{v \, \phi_b \, N} \,.
\label{eq:varepsilon}
\end{equation}
In order for $\varepsilon$ to be small, the bulk polymer density $v
\phi_b\!\gg\!2/N$, i.e.~the expansion is expected to fail for very
dilute polymer solutions. For concentrated polymer
solutions, where the bulk polymer density is much higher than the
overlap concentration, $v \phi_b\!\gg\!1/\sqrt{N}$, this type of
mean-field theory is also expected to break down. Our model is
therefore most relevant for \emph{semi-dilute} polymer
solutions.

The free energy in eq \ref{eq:Semenov_2} is written as:
\begin{equation}
\tilde{F}[f_0] \equiv \frac{2}{v \, \xi_b \, \phi_b^2} \, \frac{F[f_0]}{A \, k_{\rm B} T} = 
\int\limits_{0}^{\infty} \!\! dx \left[ (f_0^{\prime})^2 + g(f_0) \, \right]
- \frac{1}{\tilde{d}} \, f_{0,w}^2 \,,
\label{eq:rescaled_free_energy}
\end{equation}
where $g(f_0)$ is given by
\begin{equation}
g(f_0) = (f_0^2 - 1)^2 + \varepsilon \, (f_0 - 1)^2 \,.
\label{eq:rescaled_G_phi}
\end{equation}
The Euler-Lagrange equation to the free energy in
eq \ref{eq:rescaled_free_energy} reads
\begin{equation}
f_0^{\prime\prime}(x) = \frac{1}{2} \, g^{\prime}(f_0) \hspace*{30pt}
\Longrightarrow \hspace*{30pt} f_0^{\prime}(x)^2 =  g(f_0) \,,
\label{eq:rescaled_EL_1}
\end{equation}
which gives
\begin{equation}
f_0^{\prime}(x) = - \sqrt{g(f_0)} = - (f_0 - 1) \,
(f_0^2 + 2 f_0 + 1 + \varepsilon)^{1/2} \,.
\label{eq:rescaled_EL_2}
\end{equation}
As boundary condition to the above first order differential equation,
we have that $f^{\prime}_{0,w}\!=\!-f_{0,w}/\tilde{d}$. The initial value
$f_{0,w}\!\equiv\!f_0(0)$ is thus obtained by solving the following
algebraic equation
\begin{equation}
\frac{1}{\tilde{d}} f_{0,w} = (f_{0,w} - 1) \,
(f_{0,w}^2 + 2 f_{0,w} + 1 + \varepsilon)^{1/2} \,.
\label{eq:rescaled_BC_wall}
\end{equation}
The profile obtained as an explicit solution to the differential
equation reads:
\begin{equation}
f_0(x) = 1 + \frac{8 \, (\varepsilon + 4)}
{16 \, \exp(\beta (x + x_w)) - 16 - \varepsilon \, \exp(- \beta (x + x_w))} \,,
\label{eq:profile}
\end{equation}
where we have defined
\begin{eqnarray}
\exp(\beta x_w)  &\equiv& \frac{2 + 2 \, f_{0,w} + \varepsilon + \alpha \beta}
                          {4 \, (f_{0,w} - 1)} \,, \nonumber \\
\alpha           &\equiv& (f_{0,w}^2 + 2 f_{0,w} + 1 + \varepsilon)^{1/2} \,, \nonumber \\
\beta            &\equiv& (\varepsilon + 4)^{1/2} \,.
\label{eq:x0_alpha_beta}
\end{eqnarray}
This analytical result for the order parameter profile was derived
without making any further approximations. However, it should be kept
in mind that the free energy in eq \ref{eq:Semenov_2} captures only
the leading order correction (in $1/N$) to ground state dominance. The
result is that only the two leading terms in an expansion in
$\varepsilon$ are physically relevant: $f_0(x)\!=\!f_{0,0}(x) +
\varepsilon \, f_{0,1}(x) + \ldots$, with
\begin{eqnarray}
f_{0,0}(x) &=& \frac{1}{\tanh(x+x_0)} \,, \\
f_{0,1}(x) &=& - \frac{(5 \, e^{4 x_0} - 8 \, e^{2 x_0} + 4 - e^{-4 x_0})}
{32 \, (e^{4 x_0}+1) \, \sinh^2(x + x_0)}
- \frac{(4 x + 4 \, e^{- 2 x - 2 x_0} - e^{- 4 x - 4 x_0} - 3)}
{32 \, \sinh^2(x + x_0)} \,, \nonumber 
\end{eqnarray}
where we have defined $x_0\!\equiv\!(1/2) \, {\rm
arcsinh}(2\tilde{d})$.  These expressions can be used to analyze the
limiting behavior of the segment density profile for strong and weak
adsorption. For weak adsorption, we have
\begin{equation}
\frac{\phi(z)}{\phi_b} = 1 + \frac{1}{\tilde{d}} \,
\left[ e^{- 2 z/\xi_b} - \frac{\varepsilon}{8} \, (1 + \frac{2 z}{\xi_b}) \, e^{- 2 z/\xi_b}
+ \ldots \, \right] \,.
\end{equation}
In the case of strong adsorption, one finds the characteristic
$1/z^2$-behavior \cite{de_Gennes} of the segment density profile
close to the solid surface
\begin{equation}
\frac{\phi(z)}{\phi_b} = \frac{\xi_b^2}{z^2} \,.
\end{equation}
This result is unaffected by the finite chain length corrections which
implies that for strong adsorption, the segment density profile in the
immediate vicinity of the wall is unaffected by the length of the
polymer chain.

The surface tension is given by the surface free energy with the
order parameter profile, as determined by the Euler-Lagrange equation,
inserted into it:
\begin{equation}
\tilde{\sigma} \equiv \frac{2}{v \, \xi_b \, \phi_b^2} \,
\frac{\sigma}{k_{\rm B} T}
= 2 \int\limits_{0}^{\infty} \!\! dx \left[ f_0^{\prime}(x)^2 \right]
- \frac{1}{\tilde{d}} \, f_{0,w}^2 \,.
\label{eq:rescaled_sigma_int_dx}
\end{equation}
The surface tension can be conveniently rewritten as an integral
over $f$ \cite{RW}:
\begin{equation}
\tilde{\sigma} = 2 \int\limits_{1}^{f_{0,w}} \!\! df_0 \, \sqrt{g(f_0)}
- \frac{1}{\tilde{d}} \, f_{0,w}^2 \,.
\label{eq:rescaled_sigma_int_df}
\end{equation}
The value of the surface tension is thus calculated either by
inserting the density profile in eq \ref{eq:profile} back into
eq \ref{eq:rescaled_sigma_int_dx}, or by direct evaluation of the
density integral formula above. In either case, the result is:
\begin{equation}
\tilde{\sigma} = \frac{2}{3} \, \alpha \, f_{0,w} \, (f_{0,w} - 1)
+ \frac{2}{3} \, (2 - \varepsilon) \, (\beta - \alpha)
- 2 \varepsilon \, \ln \left( \frac{1 + f_{0,w} + \alpha}{2 + \beta} \right)
- \frac{1}{\tilde{d}} f_{0,w}^2 \,.
\label{eq:rescaled_sigma_result}
\end{equation}
This analytical result for the surface tension was derived without
making any further approximations.  However, it should again be kept
in mind that only the two leading terms in an expansion in
$\varepsilon$ are physically relevant:
\begin{equation}
\tilde{\sigma} = \tilde{\sigma}_0 + \varepsilon \,\tilde{\sigma}_1 + \ldots \,.
\label{eq:rescaled_sigma_expansion}
\end{equation}
Expanding eq \ref{eq:rescaled_sigma_result} in $\varepsilon$
gives as explicit expressions:
\begin{eqnarray}
\tilde{\sigma}_0 &=& \frac{4}{3} - \frac{4}{3 \, \tanh(x_0)} 
- \frac{\cosh(x_0)}{3 \, \sinh^3(x_0)} \,, \nonumber \\
\tilde{\sigma}_1 &=& \frac{1}{\tanh(x_0)} - 1  + 2 \, \ln(1-e^{-2 x_0}) \,.
\label{eq:rescaled_sigma_expansion_coeff}
\end{eqnarray}
The result for $\tilde{\sigma}_0$ was first derived in
ref.\cite{Skau}.  In Figure 2, the surface tension is plotted as a
function of chain length at fixed $\tilde{d}\!=\!1$ and for various
values of $v \phi_b$. The solid line is the full result in
eq \ref{eq:rescaled_sigma_result} with the dashed line the two
leading terms (eqs \ref{eq:rescaled_sigma_expansion} and
\ref{eq:rescaled_sigma_expansion_coeff}). Even though the relevant
variable is the combination $\varepsilon\!=\!2/(N \, v \, \phi_b)$, we
have chosen to plot the surface tension as a function of $N$ for
various $v \phi_b$.  The condition that $\varepsilon$ is small implies
that $N\!\gg\!1/(v \, \phi_b \, \varepsilon)$.  This limit corresponds
(for each $v \phi_b$) roughly to the point where the solid and dashed
lines deviate.  The leading order correction to the surface tension is
always positive indicating that the lowering of the surface tension
due to the presence of adsorbed polymer is \emph{reduced} when the
polymer chain becomes shorter. The magnitude of this effect is
appreciable ($\simeq\!$ 10 \%) already for large polymer chains
($N\!\simeq\!$ 1000) when the bulk density is low ($v
\phi_b\!\simeq\!$ 0.001).

It is instructive to consider the behavior of $\tilde{\sigma}_1$
in the weak adsorption limit ($d\gg\xi_b$):
\begin{equation}
\tilde{\sigma}_1 = \frac{1}{16} \frac{1}{\tilde{d}^2}
+ \frac{1}{48} \frac{1}{\tilde{d}^3} + {\cal O}(\frac{1}{\tilde{d}^4}) \,.
\end{equation}
This result for $\tilde{\sigma}_1$ can be combined with the limiting
behavior for $\tilde{\sigma}_0$.  To show this in more detail, we
subtract from the surface tension the constant contribution to
$\sigma$ that remains even when the density profile is equal to the
bulk density everywhere, $\phi(z)\!=\!\phi_b$. One can then write
\begin{equation}
\Delta \sigma \equiv \sigma + k_{\mathsf{B}} T \frac{a^2}{6}\frac{\phi_b}{d}
= - \frac{a^4}{36} \frac{k_{\mathsf{B}} T}{\mathit{v} \, \xi_b \, d^2}
\left( 1 - \frac{\varepsilon}{8} + \ldots \right) \,.
\label{eq:sigma_expansion_weak}
\end{equation}
This expression more clearly shows that the lowering of the surface
tension due to the presence of adsorbed polymer is \emph{reduced} when
the polymer chain becomes shorter. One may show \cite{preprint} that
the next term in the expansion of the surface tension in
eq \ref{eq:sigma_expansion_weak} goes as $\simeq -0.1540 \,
\varepsilon^{3/2}$.

For completeness we also give the expression for $\tilde{\sigma}_1$ in
the case of \emph{strong} polymer adsorption $(d\!\ll\!\xi_b)$:
\begin{equation}
\tilde{\sigma}_1 = \frac{1}{\tilde{d}} + \left( 2\ln(2\tilde{d}) - 1 \right)
\, \tilde{d} + {\cal O}(\tilde{d}^2) \,.
\end{equation}
%
\section{Semenov model for adsorption onto curved surfaces}

We now extend our analysis to curved surfaces. In particular, we are
interested in determining the coefficients of an expansion of the free
energy to second order in the curvature: the spontaneous radius of
curvature, $R_0$, and the rigidity constants $k$ and $\bar{k}$.  In
general geometry, the free energy functional in the Semenov model
reads:
\begin{equation}
\frac{F[\phi]}{k_{\rm B} T} = \int \!\! d\vec{r} \left[
\frac{a^2}{24} \, \frac{|\vec{\nabla}\phi(\vec{r})|^2}{\phi}
+ G(\phi) \, \right] - \frac{A}{d} \, \frac{a^2}{6} \, \phi_{w} \,,
\label{eq:Semenov_curved_free_energy}
\end{equation}
We rescale the variables as done previously, see eq \ref{eq:rescaling}.
The rescaled free energy is then given by:
\begin{equation}
\tilde{F}[f] = \frac{1}{A} \int \!\! d\vec{x}
\left[ |\vec{\nabla}f(\vec{x})|^2 + g(f) \, \right]
- \frac{1}{\tilde{d}} \, f_w^2 \,,
\label{eq:Semenov_curved_free_energy_rescaled}
\end{equation}
with the Euler-Lagrange equation and boundary condition:
\begin{equation}
\Delta f(\vec{x}) = \frac{1}{2} \, g^{\prime}(f) \,, \hspace*{50pt}
\hat{n} \cdot \vec{\nabla}f_w = - \frac{1}{\tilde{d}} f_w \,.
\label{eq:Semenov_curved_BC_wall}
\end{equation}
To derive the curvature parameters, both order parameter profile and
free energy are expanded to second order in curvature for a
spherically and cylindrically shaped surface \cite{Blokhuis_93}.  For
example, in the spherical geometry we have:
\begin{equation}
f_s(x) = f_0(x) + \frac{1}{R} \, f_{s,1}(x) + \frac{1}{R^2} \, f_{s,2}(x)
+ \ldots \,,
\label{eq:expansion_f_sphere}
\end{equation}
where $R$ is the sphere's radius.  In this expansion, the
Euler-Lagrange equation and boundary condition are given to zeroth and
first order by:
\begin{eqnarray}
f_0^{\prime\prime}(x) &=& \frac{1}{2} \, g^{\prime}(f_0) \,, \hspace*{103pt}
f_{0,w}^{\prime} = - \frac{1}{\tilde{d}} f_{0,w} \,, \nonumber \\
f_{s,1}^{\prime\prime}(x) &=& \frac{1}{2} \, g^{\prime}(f_0) \, f_{s,1}(x) 
- 2 \, f_0^{\prime}(x) \,, \hspace*{20pt}
f_{s,1,w}^{\prime} = - \frac{1}{\tilde{d}} f_{s,1,w} \,.
\label{eq:expansion_EL_BC}
\end{eqnarray}
The spontaneous radius of curvature, $R_0$, and the rigidity constants
$k$ and $\bar{k}$, are given by \cite{Clement_Joanny,Skau,Blokhuis_93}
\begin{eqnarray}
\tilde{c}_0 &\equiv& \left( \frac{2}{v \, \xi_b^2 \,
\phi_b^2 \, k_{\rm B} T} \right) \, \frac{2k}{R_0} =
- 2 \int\limits_{0}^{\infty} \!\! dx \left[ x \, f_0^{\prime}(x)^2 \right]
\,, \nonumber \\
\tilde{\bar{k}} &\equiv& \left( \frac{2}{v \, \xi_b^3 \, \phi_b^2 \, k_{\rm B} T} \right) \,
\bar{k} \hspace*{9pt} = 2 \int\limits_{0}^{\infty} \!\! dx
\left[ x^2 \, f_0^{\prime}(x)^2 \right] \,, \nonumber \\
\tilde{k} &\equiv& \left( \frac{2}{v \, \xi_b^3 \, \phi_b^2 \, k_{\rm B} T} \right) \, k
\hspace*{9pt} = - \int\limits_{0}^{\infty} \!\! dx \left[ f_0^{\prime}(x)
f_{s,1}(x) \right] \,.
\label{eq:curvature_parameters_int_dx}
\end{eqnarray}
The expressions for the spontaneous radius of curvature, $R_0$, and
the rigidity constants associated with Gaussian curvature, $\bar{k}$,
can be rewritten as integrals over $f$, similar to the expression for
the surface tension in eq \ref{eq:rescaled_sigma_int_dx}:
\begin{eqnarray}
\tilde{c}_0 &=&
- 2 \int\limits_{1}^{f_w} \!\! df \, \sqrt{g(f)} \,
\int\limits_{f}^{f_w} \!\! df^{\prime} \frac{1}{\sqrt{g(f^{\prime})}}
\,, \nonumber \\
\tilde{\bar{k}} &=& 2 \int\limits_{1}^{f_w} \!\! df \sqrt{g(f)} \, \left[
\int\limits_{f}^{f_w} \!\! df^{\prime} \frac{1}{\sqrt{g(f^{\prime})}}
\, \right]^2 \,.
\label{eq:curvature_parameters_int_df}
\end{eqnarray}
One may show that the bending rigidity constant, $k$, is given by
\begin{eqnarray}
\tilde{k} &=& \frac{4 \, \tilde{d}}{\sqrt{g(f_w)} \,\,
[2 \, \sqrt{g(f_w)} - \tilde{d} \, g^{\prime}(f_w)]} \,
\left[ \int\limits_{1}^{f_w} \!\! df \, \sqrt{g(f)} \, \right]^{\!2} \nonumber \\
&& -2 \int\limits_{1}^{f_w} \!\! df \, \sqrt{g(f)} \,
\int\limits_{f}^{f_w} \!\! df^{\prime} \frac{1}{\sqrt{g(f^{\prime})^3}}
\int\limits_{1}^{f^{\prime}} \!\! df^{\prime\prime} \sqrt{g(f^{\prime\prime})} \,.
\label{eq:rigidity_constant}
\end{eqnarray}
Explicit results for the value of $\tilde{c}_0$, $\tilde{\bar{k}}$,
and $\tilde{k}$ is obtained by inserting the functional form for $g(f)$
(eq \ref{eq:rescaled_G_phi}) into
eqs \ref{eq:curvature_parameters_int_df} and
\ref{eq:rigidity_constant}, and evaluating the resulting integrals
numerically. Typical results for the various curvature parameters as a
function of chain length are shown as the solid lines in Figure
3. Again, it should be kept in mind that the physical relevance of
these expressions for the curvature parameters is restricted to the
first two terms in an expansion in $\varepsilon$:
\begin{eqnarray}
\tilde{c}_0     &=& \tilde{c}_{0,0} + \varepsilon \, \tilde{c}_{0,1} + \ldots \,, \nonumber \\
\tilde{\bar{k}} &=& \tilde{\bar{k}}_0 + \varepsilon \, \tilde{\bar{k}}_1 + \ldots \,, \nonumber \\
\tilde{k}       &=& \tilde{k}_0 + \varepsilon \, \tilde{k}_1 + \ldots \,.
\label{eq:rescaled_curvature_parameters_expansion}
\end{eqnarray}
An explicit calculation gives \cite{Skau}:
\begin{eqnarray}
\tilde{c}_{0,0}   &=& - \frac{1}{3 \, \sinh^2(x_0)} - \frac{4}{3} \, \ln(1-e^{-2 x_0}) \,, \nonumber \\
\tilde{\bar{k}}_0 &=& - \frac{2}{3} + \frac{2}{3 \, \tanh(x_0)} 
                      - \frac{4}{3} \, {\rm dilog}(1-e^{-2 x_0}) \,, \nonumber \\
\tilde{k}_0       &=& - \,
\frac{(27 e^{2 x_0} + 5 - e^{-2 x_0} + e^{-4 x_0})}{18 \, (e^{2 x_0}-1) \, (e^{4 x_0}+1)} \,,
\label{eq:rescaledcurvatureparameters_expansion_0}
\end{eqnarray}
and
\begin{eqnarray}
\tilde{c}_{0,1} &=& - \,
\frac{(4 e^{2 x_0} - 7 + e^{-4 x_0})}{12 \, (e^{4 x_0} + 1)}
+ {\rm dilog}(1-e^{-2 x_0}) + \frac{2}{3} \, \ln(1-e^{-2 x_0}) \,, \nonumber \\
\tilde{\bar{k}}_1 &=& - \, \frac{(3 e^{2 x_0} - e^{-2 x_0})}{12 \, (e^{4 x_0}+1)}
- \, \frac{(5 e^{4 x_0} - 8 e^{2 x_0} + 4 - e^{-4 x_0})}
{12 \, (e^{4 x_0}+1)} \, \ln(1-e^{-2 x_0}) \nonumber \\
&& + \frac{5}{6} \, {\rm dilog}(1-e^{-2 x_0})
- {\rm Li_{3}}(e^{-2 x_0}) \,, \nonumber \\
\tilde{k}_1       &=& - \, \frac{1}{288 \, (e^{4x_0}+1)^3} \left[
456 \, e^{10 x_0} - 354 \, e^{8 x_0} + 1272 \, e^{6 x_0} - 281 \, e^{4 x_0} \right.
+ 792 \, e^{2 x_0} \nonumber \\
&& \hspace*{95pt} \left. - 143 + 168 \, e^{-2 x_0} + 37 \, e^{-4 x_0} - 3 \, e^{-8 x_0} \right] \\
&& - \, \frac{(13 e^{4 x_0} - 16 e^{2 x_0} + 4 - e^{-4 x_0})}{12 \, (e^{4 x_0}+1)} \,
\ln(1-e^{-2 x_0}) + \frac{1}{2} \, {\rm dilog}(1-e^{-2 x_0}) \nonumber \,,
 \label{eq:rescaledcurvatureparameters_expansion_1}
\end{eqnarray}
where the polylogs ${\rm Li_{n}}(z)$ and ${\rm dilog(z)}$ are defined
in the Appendix.

The two leading contributions
eq \ref{eq:rescaledcurvatureparameters_expansion_0} with eq 3.11
are shown as the dashed curves in Figure 3.  Again, the validity limit
of the expansion in $\varepsilon$ corresponds (for each $v \phi_b$)
roughly to the point where the solid and dashed lines deviate.  The
curves in Figure 3 show that the leading order correction to the curvature
parameters is significant already for large polymer chains, especially
when the bulk polymer density is low. Furthermore, when the polymer
chain becomes shorter, the effect of polymer adsorption on the
curvature parameters is reduced.  However, this is not a general
result and it is only valid when the adsorption strength is
sufficiently weak. This is shown in Figure 4 where we have plotted the
leading order corrections to the curvature parameters as a function of
the adsorption strength.  For large $\tilde{d}$, i.e.~weak adsorption,
the leading order corrections are such that they are opposite in sign
to the ground state dominance results ($\tilde{c}_{0,1}$ and
$\tilde{k}_1$ are positive and $\tilde{\bar{k}}_1$ is negative); for
small $\tilde{d}$, i.e.~strong adsorption, they \emph{enhance} the
effect.

These results are further demonstrated by considering the limiting
behavior.  In the weak adsorption limit ($d\gg\xi_b$) one finds:
\begin{eqnarray}
\tilde{c}_{0,1} &=& \frac{1}{32} \frac{1}{\tilde{d}^2} + \frac{1}{288} \frac{1}{\tilde{d}^3}
+ {\cal O}(\frac{1}{\tilde{d}^4}) \,, \nonumber \\
\tilde{\bar{k}}_1 &=& - \frac{3}{128} \frac{1}{\tilde{d}^2} + \frac{1}{576} \frac{1}{\tilde{d}^3}
+ {\cal O}(\frac{1}{\tilde{d}^4}) \,, \nonumber \\
\tilde{k}_1 &=& \frac{9}{256} \frac{1}{\tilde{d}^2} - \frac{1}{96} \frac{1}{\tilde{d}^3}
+ {\cal O}(\frac{1}{\tilde{d}^4}) \,.
\label{eq:rescaledcurvatureparameters_expansion_1weak}
\end{eqnarray}
The leading order corrections all scale as $1/\tilde{d}^2$ just as the
ground state dominance results for $\tilde{c}_{0,0}$,
$\tilde{\bar{k}}_0$, and $\tilde{k}_0$ \cite{Skau}.  We can therefore
write:
\begin{eqnarray}
\frac{2k}{R_0} &=& - \frac{a^4}{144}\frac{k_{\mathsf{B}} T}{\mathit{v} \,d^2}\,
              \bigl( 1 - \frac{\varepsilon}{4} + \ldots \bigr) \,, \nonumber \\
\bar{k} &=& \frac{a^4}{288} \frac{k_{\mathsf{B}} T \,\xi_b}{\mathit{v} \,d^2}\,
                  \bigl( 1 - \frac{3 \varepsilon}{8} + \ldots\bigr)\,,\nonumber \\
k &=& - \frac{a^4}{192} \frac{k_{\mathsf{B}} T \, \xi_b}{\mathit{v} \, d^2} \,
                  \bigl( 1 - \frac{3 \varepsilon}{8} + \ldots \bigr) \,.
\end{eqnarray}
These expressions show more clearly that in the weak adsorption limit,
the influence on the curvature parameters caused by the presence of
adsorbed polymer is reduced when the polymer chain becomes shorter.

For strong adsorption $(d\!\ll\!\xi_b)$ one has:
\begin{eqnarray}
\tilde{c}_{0,1} &=& \frac{2}{3} \, \ln(2\tilde{d}) + \frac{\pi^2}{6} + \frac{1}{12}
+ \left( 2\ln(2\tilde{d}) - 3 \right) \, \tilde{d} + {\cal O}(\tilde{d}^2) \,, \nonumber \\
\tilde{\bar{k}}_1 &=& \frac{5 \pi^2}{36} - \frac{1}{12} - \zeta(3)
+ \left( \frac{4}{3} \ln(2\tilde{d}) +\frac{\pi^2}{3} - \frac{11}{6} \right) \, \tilde{d}
+ {\cal O}(\tilde{d}^2) \,, \nonumber \\
\tilde{k}_1 &=& - \frac{27}{32} + \frac{\pi^2}{12} + {\cal O}(\tilde{d}^2) \,,
\label{eq:rescaled_curvature_parameters_expansion_1_strong}
\end{eqnarray}
where the zeta function $\zeta(z)$ is defined in the Appendix.

These results show that when the adsorption is sufficiently strong,
one finds the somewhat surprising result that shortening the polymer
chain \emph{enhances} the effect on the curvature parameters induced by
adsorbing polymer.

\section{Discussion}

In this article we have investigated the chain length dependence of
the curvature properties of polymer adsorbed surfaces. Our
calculations are done in the context of the Semenov model
\cite{Semenov_96_review} for the free energy which captures the
leading correction to the free energy in an expansion in
$1/N$. Analytic expressions are derived for the leading corrections to
the surface tension and curvature parameters.

For an infinite chain, it is well-established that the ground state
dominance model predicts that the addition of polymer reduces the
surface tension, induces a spontaneous curvature towards the polymers,
gives a positive contribution to the Gaussian rigidity, and reduces
the value of the bending rigidity. We find that if the adsorption
strength is sufficiently low, the leading correction due to the
finiteness of the polymer chain to the surface tension and curvature
parameters is to \emph{reduce} the effect of polymer adsorption on
these coefficients. The magnitude of this effect may be considerable
already for long polymer chains, especially when the bulk polymer
density is low.

For strong adsorption, we find that the leading contribution to the
curvature parameter \emph{changes sign}. This means that shortening the
polymer chains now \emph{enhances} the effect polymers have on the
curvature parameters. This seems a bit surprising since it is
expected that for very short chains, the influence of these chains
becomes less pronounced. Still, our result is not necessarily in contradiction
with this expectation since for small $N$ \emph{all} the terms in
the expansion in $1/N$ are important.

One would like to understand the \emph{physical reason} behind this
``enhancement effect''. Mathematically, the effect is traced back to
the non-monotonic behavior of the leading correction to the segment
density profile $f_{0,1}(x)$, which is directly linked to the leading
corrections to the spontaneous curvature ($\tilde{c}_{0,1}$) and the
Gaussian rigidity ($\tilde{\bar{k}}_1$).  (An equivalent argument can
be made for the bending rigidity ($\tilde{k}_1$) which is determined
by the behavior of $f_{1,1}(x)$.)

In figure 5, we have plotted $f_{0,1}(x)$ for various values of the
adsorption strength. For weak adsorption, $f_{0,1}(x)$ monotonically
increases to zero at large distances indicating that short chains
adsorb less. For strong adsorption, two regions may be identified: in
the vicinity of the wall the segment density profile is dominated by
polymer `loops'. In this region $f_{0,1}(x)$ is close to zero
indicating that the polymer density is independent of chain length. At
somewhat larger distances from the wall the profile is dominated by
the polymer `tails' [2, 24]. Here $f_{0,1}(x)$ goes through a minimum
and then monotonically increases to zero at large distances.  One may
therefore conclude that, although the total adsorption is less when
the chain length is reduced, the segregation between a region
dominated by loops and a region dominated by tails is more pronounced
leading to steeper gradients in the polymer segment density profile.
This is then ultimately responsible for an enhancement of the polymer
contribution to the curvature parameters in the case of strong
adsorption.

We end with a discussion of the limitations of our theoretical
treatment.  Corrections to the curvature parameters were calculated to
leading order in $1/N$. We showed that the leading correction to the
segment density profile gives a modified density profile that varies
on the scale of the bulk correlation length, as does the ground state
dominance profile. One expects \cite{Semenov_96_review, preprint } higher order
corrections (${\cal O}(1/N^{3/2})$) to give modifications to the
segment density profile on the scale of the polymer's radius of
gyration.  Although these higher order corrections dominate the
segment density profile when $z\!\gg\!\xi_b$, they remain subdominant
when it concerns the curvature parameters since these are expressed in
terms of integrals over the entire region of inhomogeneity.

A subtle point concerns the dimensional dependence of our
analysis. Since all our results are derived within mean-field theory,
it is implied that the critical exponents are those of a polymer
system embedded in 4 dimensions. In our derivation of the curvature
parameters we explicitly consider the geometry to be that of a
2-dimensional surface curved in 3-dimensional space, which is arguably
inconsistent with the mean-field assumption. It seems hard to avoid
such an inconsistency, however, and it is a critique which applies to
all mean-field calculations of curvature properties.

The interaction with the solid surface is modeled by the
extrapolation length $d$ giving an effective interaction. Details of
the structure of the wall (e.g.~surface roughness, surface
heterogeneity) and the full wall-polymer interaction potential are
therefore not explicitly considered. It is expected that these
factors influence the polymer segment density in the very vicinity of
the wall. In the present calculations, the scale of the segment
density profile is set by the bulk correlation length. One expects
that the detailed structure of the wall is unimportant as long as the
bulk correlation length is much larger than the microscopic length
scale associated with this structure.

Formally, in the calculation presented the surface is taken to be that
of an undeformable, attractive solid wall so that the main application
of our results to experiments is for, say, polymer adsorption onto a
colloidal particle. However, our results should also be of interest to
the description of polymer adsorption on more flexible surfaces such
as membranes \cite{Stavans}, for which the essential physics is
captured by the Helfrich free energy.

\appendix

\section{Mathematical Functions}

As a reference, we provide definitions of the special functions
used.

\noindent
The Polylogarithm function is defined as:
\begin{equation}
{\rm Li_{n}}(z) \equiv \sum\limits_{k=1}^{\infty} \frac{z^k}{k^n} \,.
\end{equation}
The Dilogarithm function is defined as:
\begin{equation}
{\rm dilog}(z) \equiv {\rm Li_{2}}(z) \equiv \int\limits_1^z \!\! dt \frac{\ln t}{(1-t)} \,.
\end{equation}
The Riemann zeta function is defined as:
\begin{eqnarray}
\zeta(n) &\equiv& {\rm Li_{n}}(1) \equiv \sum\limits_{k=1}^{\infty} \frac{1}{k^n} \,,\\
\zeta(3) &\simeq&  1.202056903\ldots
\end{eqnarray}

\vskip 20pt
\noindent
{\Large\bf Acknowledgments}
\vskip 5pt
\noindent
We would like to thank J. Bonet-Avalos and A. Johner 
for helpful discussions and comments.

\newpage

\begin{figure}
  \centering
   \includegraphics[width=100mm]{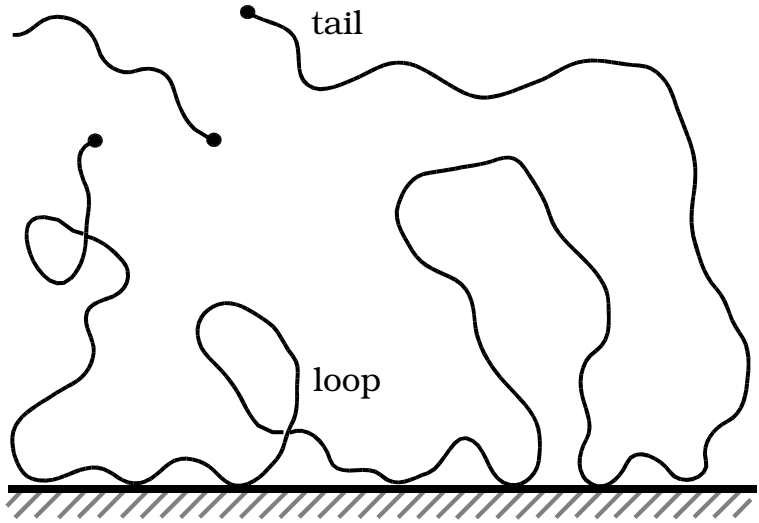}
   \caption{Sketch of a polymer chain adsorbed to a planar, solid
surface located at $z\!=\!0$. The polymer chain consists of a series
of `loops', with both ends adsorbed to the substrate, and with two
`tails', having one end free.}
\end{figure}

\begin{figure}
  \centering
    \includegraphics[width=140mm]{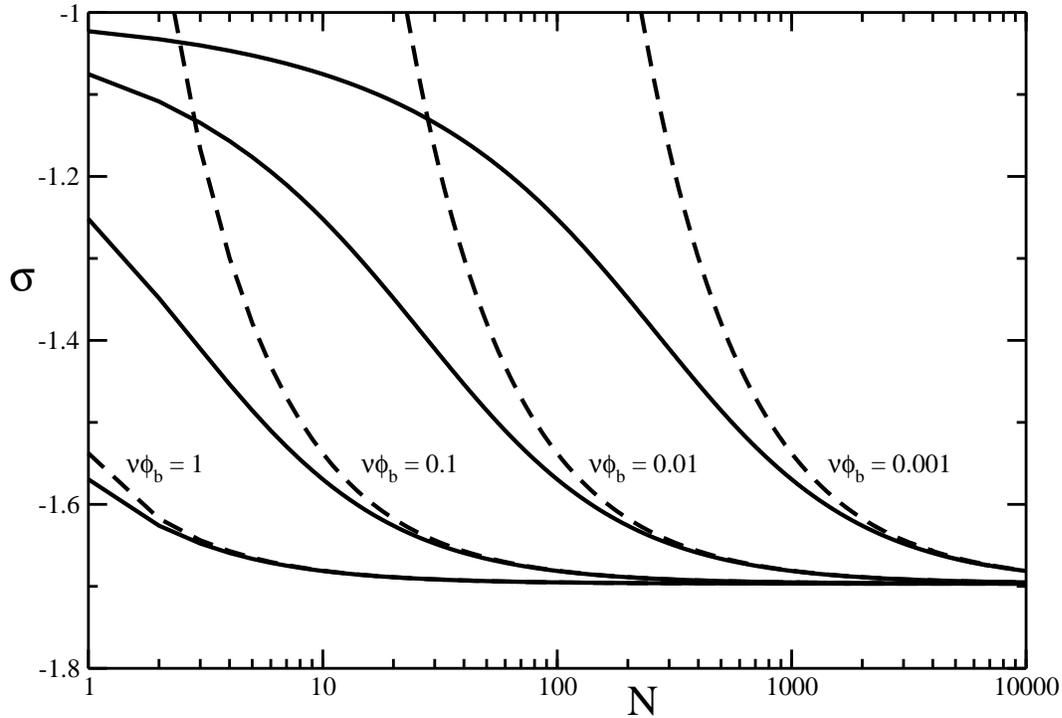} 
   \caption{Surface tension $\tilde{\sigma}$ as a function of chain
length. The adsorption strength $\tilde{d}\!=\!1$; the bulk polymer
density ranges from $v \phi_b\!=\!0.001$ to $v \phi_b\!=\!1$. The
solid line is eq \ref{eq:rescaled_sigma_result}; the dashed line is eq
\ref{eq:rescaled_sigma_expansion} combined with eq
\ref{eq:rescaled_sigma_expansion_coeff}.}
\end{figure}
 
\begin{figure}
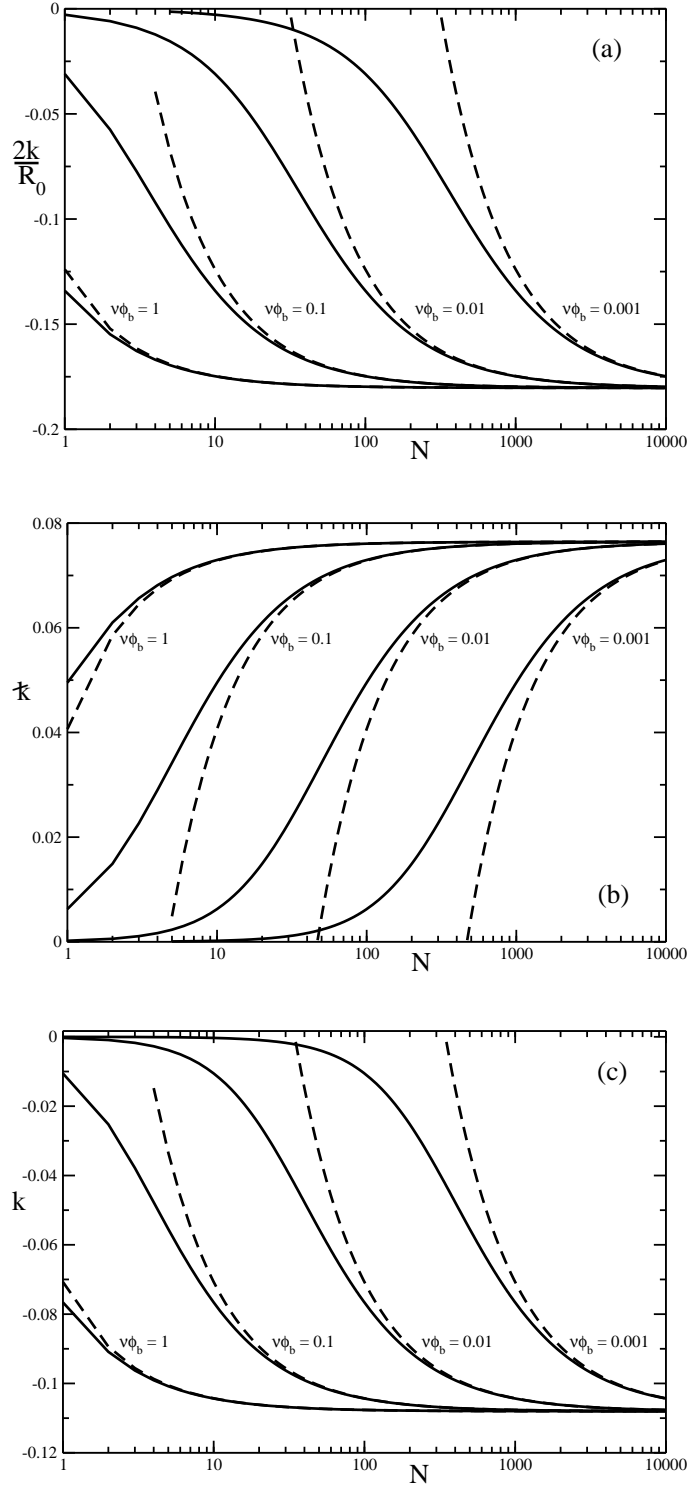

     \centering
     \subfigure{
        \includegraphics[width=90mm]{Fig3a.eps}
           \label{Fig:3a} }\\
     \subfigure{
        \includegraphics[width=90mm]{Fig3b.eps}
           \label{Fig:3b} }\\
      \subfigure{ 
	\includegraphics[width=90mm]{Fig3c.eps}
           \label{Fig:3c} }\\
     \caption{\small The curvature parameters as a function of chain length;
(a) is the spontaneous curvature $\tilde{c_0}$, (b) is the Gaussian
rigidity $\tilde{\bar{k}}$, (c) is the bending rigidity
$\tilde{k}$. The adsorption strength $\tilde{d}\!=\!1$; the bulk
polymer density ranges from $v \phi_b\!=\!0.001$ to $v
\phi_b\!=\!1$. The solid lines are numerical evaluations of the
integral expressions in eqs \ref{eq:curvature_parameters_int_df} and
\ref{eq:rigidity_constant}. The dashed lines are the two leading
contributions, eq \ref{eq:rescaledcurvatureparameters_expansion_0}
combined with eq 3.11.}
     \label{fig:2858multifig}
\end{figure}

\begin{figure}
 \centering
  \includegraphics[width=130mm]{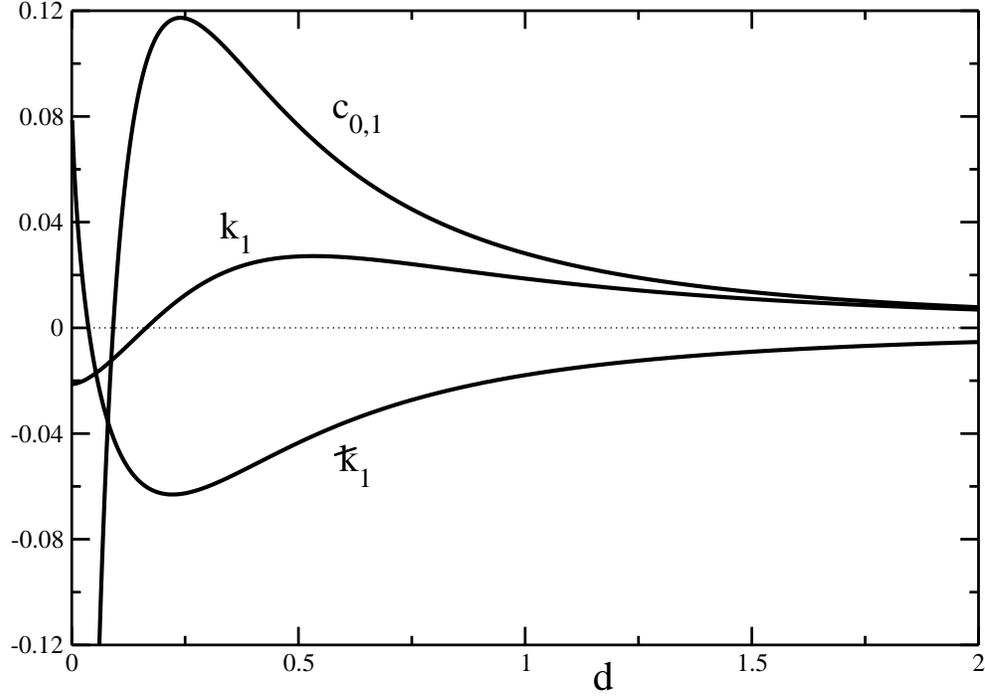} 
   \caption{Leading order corrections to the curvature parameters:
$\tilde{c}_{0,1}$, $\tilde{\bar{k}}_1$, and $\tilde{k}_1$, plotted as
a function of the adsorption strength (strong adsorption when
$\tilde{d} \ll 1$).}
\end{figure}

\begin{figure}
 \centering
  \includegraphics[width=130mm]{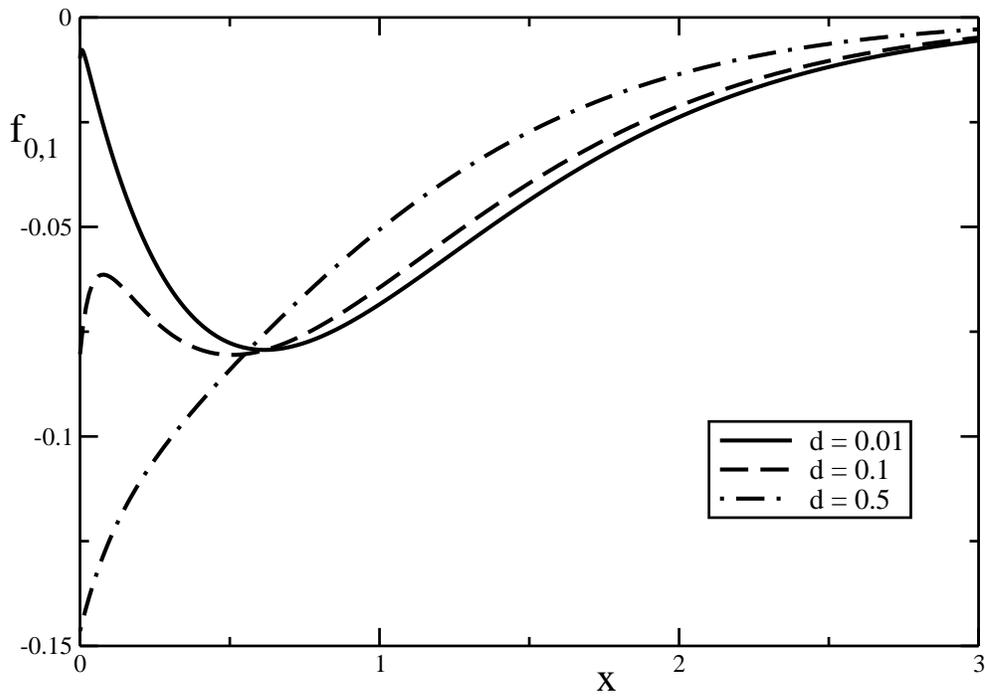} 
   \caption{The order parameter profile $f_{0,1}(x)$ as a function of
$x$ for various values of the adsorption strength; solid line:
$\tilde{d}\!=\!0.01$, dashed line: $\tilde{d}\!=\!0.1$, dot-dashed
line: $\tilde{d}\!=\!0.5$.}
\end{figure}


\newpage
\noindent
\begin{thebibliography}{}

\bibitem{de_Gennes}
de Gennes, P.G. \emph{Scaling Concepts in Polymer Physics};
Cornell University Press: Ithaca, 1979.

\bibitem{Fleer_book} Fleer, G. J.; Cohen Stuart, M. A.; Scheutjens, J. M. H. M.; Cosgrove, T.; Vincent, B. \emph{Polymers at Interfaces}; Chapman and Hall: London, 1993.

\bibitem{Eisenriegler_book} Eisenriegler, E. \emph{Polymers near Interfaces};
World Scientific: Singapore, 1993.

\bibitem{Odijk} Odijk, T. \emph{Macromolecules} \textbf{1996},
\emph{29}, 1842. Odijk, T. \emph{Physica A} \textbf{2000}, \emph{278}, 347.

\bibitem{Johner_01} Johner, A.; Joanny, J.-F.; Diez Orrite, S.;
Bonet-Avalos, J. \emph{Europhys. Lett.} \textbf{2001}, \emph{56}, 549.

\bibitem{Lekkerkerker}  Tuinier, R.; Vliegenthart, G.A.; Lekkerkerker, H.N.W.
\emph{J. Chem. Phys.} \textbf{2000}, \emph{113}, 10768.

\bibitem{Eisenriegler_Dietrich}  Hanke, A.; Eisenriegler, E.; Dietrich, S.
\emph{Phys. Rev. E} \textbf{1999}, \emph{59}, 6853.

\bibitem{Maasen}  Maasen, R.; Eisenriegler, E.;  Bringer, A.
\emph{J. Chem. Phys.} \textbf{2001}, \emph{115}, 5292.

\bibitem{Louis}  Louis, A.A.; Bolhuis, P.G.; Meijer, E.J.;
Hansen, J.P. \emph{J. Chem. Phys.} \textbf{2002}, \emph{117}, 1893.

\bibitem{Helfrich_Canham} Helfrich, W. \emph{Z. Naturforsch.}
 \textbf{1973}, \emph{28c}, 693.  Canham, P.B. \emph{J. Theor. Biol.}
 \textbf{1970}, \emph{26}, 61.

\bibitem{Ji_Hone} Ji, H.; Hone, D. \emph{Macromolecules} \textbf{1988},
\emph{21}, 2600.

\bibitem{Brooks_Marques_Cates} Brooks, J.T.; Marques, C.M.; Cates,
 M.E. \emph{J. Phys. II} \textbf{1991}, \emph{1}, 673. Brooks, J.T.;
 Marques, C.M.; Cates, M.E. \emph{Europhys. Lett.}  \textbf{1991},
 \emph{14}, 713 .

\bibitem{Podgornik} Podgornik, R. \emph{Europhys. Lett.}
\textbf{1993}, \emph{21}, 245 .

\bibitem{Clement_Joanny}  Clement, F.;  Joanny, J.-F.
\emph{J. Phys. II} \textbf{1997}, \emph{7}, 973.

\bibitem{Skau}  Skau, K.I.;  Blokhuis, E.M. \emph{Eur. Phys. J. E.}
\textbf{2002}, \emph{7}, 13.

\bibitem{Edwards} Edwards, S.F. \emph{Proc. Phys. Soc.}
\textbf{1965}, \emph{85}, 613. Edwards, S.F. \textbf{1966},
\emph{88}, 265.

\bibitem{Lifshitz} Lifshitz, I.M. \emph{Soviet Phys. JETP}
\textbf{1969}, \emph{28}, 1280.  Lifshitz, I.M.; Grosberg, A.Yu.;
Khoklov, A.R. \emph{Rev. Mod. Phys.} \textbf{1978}, \emph{50}, 684.

\bibitem{Simulations} de Joannis, J.; Ballamudi, R.K.; Park, C.-W.;
Thomatos, J.; Bitsanis, I.A. \emph{Europhys. Lett.} \textbf{2001},
\emph{56}, 200.  de Joannis, J.; Park, C.-W.; Thomatos, J.; Bitsanis,
I.A. \emph{Langmuir} \textbf{2001}, \emph{17}, 69.

\bibitem{Semenov_95} Semenov, A.N.; Joanny,
J.-F. \emph{Europhys. Lett.} \textbf{1995}, \emph{29}, 279.

\bibitem{Semenov_96}
Semenov, A.N.;  Bonet-Avalos, J.;  Johner, A.;  Joanny, J.-F.
\emph{Macromolecules} \textbf{1996}, \emph{29}, 2179.

\bibitem{Johner_96} Johner, A.; Bonet-Avalos, J.; van der Linden,
 C.C.; Semenov, A.N.; Joanny, J.-F. \emph{Macromolecules}
 \textbf{1996}, \emph{29}, 3629.

\bibitem{Bonet_96}  Bonet-Avalos, J.;  Joanny, J.-F.;  Johner, A.;
Semenov, A.N. \emph{Europhys. Lett.} \textbf{1996}, \emph{35}, 97.
Semenov, A.N.;  Joanny, J.-F.;  Johner, A.;   Bonet-Avalos, J.
\emph{Macromolecules} \textbf{1997}, \emph{30}, 1479.

\bibitem{Semenov_96_review} Semenov, A.N. \emph{J. Phys. II}
 \textbf{1996}, \emph{6}, 1759.

\bibitem{Semenov_98} Semenov, A.N.; Joanny, J.-F.; Johner, A.
 \emph{Polymer Adsorption: mean-field theory and ground state
 dominance approximation}; In: \emph{Theoretical and Mathematical Models in
 Polymer Research};  Grosberg, A., Ed.; Academic Press: San Diego,
 1998; p.37.

\bibitem{de_Gennes_82}
de Gennes, P.G. \emph{Macromolecules} \textbf{1982}, \emph{15}, 492.

\bibitem{RW}
Rowlinson, J.S.;  Widom, B. \emph{Molecular Theory of Capillarity};
Clarendon: Oxford, 1984.

\bibitem{preprint}
Blokhuis, E.M.; Skau, K.I.;  Bonet-Avalos, J. ``Free energy formalism for 
polymer adsorption: self consistent field theory for weak adsorption.''
\emph{preprint}.

\bibitem{Blokhuis_93}
 Blokhuis, E.M.;  Bedeaux, D. \emph{Mol. Phys.} \textbf{1993}, \emph{80}, 705.

\bibitem{Stavans} Frette, V.; Tsafrir, I.; Guedeau-Boudeville, M.-A.;
Jullien, L.; Kandel, D.; Stavans, J. \emph{Phys. Rev. Lett.}
\textbf{1999}, \emph{83}, 2465.  Tsafrir, I.; Guedeau-Boudeville,
M.-A.; Kandel, D.; Stavans, J. \emph{Phys. Rev. E} \textbf{2001},
\emph{63}, 031603 .

\end{thebibliography}
\end{document}